\documentclass{midl} % Include author names
\usepackage{verbatim}
\jmlrproceedings{MIDL}{Medical Imaging with Deep Learning}
\jmlrpages{}
\jmlryear{2019}
\jmlrworkshop{MIDL 2019 -- Extended Abstract Track}

\title[SPECT Denoising using GAN]{Low Dose SPECT Image Denoising Using a Generative Adversarial Network}

\midlauthor{\Name{Qi Zhang\nametag{$^{1}$}} \Email{mb85402@um.edu.mo} \\
\addr $^{1}$ Department of Computer and Information Science, Faculty of Science and Technology, University of Macau, Macau SAR, People’s Republic of China \\
\Name{Jingzhang Sun\nametag{$^{2}$}} \Email{yb87437@um.edu.mo}\\
\addr $^{2}$ Department of Electrical and Computer Engineering, Faculty of Science and Technology, University of Macau, Macau SAR, People’s Republic of China \\
\Name{Greta S. P. Mok\nametag{$^{2,3}$}} \Email{gretamok@um.edu.mo}\\
\addr $^{3}$ Faculty of Health Sciences, University of Macau, Macau SAR, People’s Republic of China \AND
}

\begin{document}

\maketitle
\vspace{-15pt}
\section{Introduction}
Single-photon emission computed tomography (SPECT) is an in vivo functional imaging technique that uses gamma cameras to detect molecular-level activities of patients’ tissues generally through injection of the radio-labelled pharmaceuticals. The image noise level and resolution of SPECT images are often poor, due to the limited number of detected counts and various physical degradation factors during SPECT acquisition \cite{garcia2012physical}. This problem has considerably affected lesion detection, clinical diagnosis and treatment.\\
Recently generative adversarial networks (GAN) have been proved successfully in numerous computer vision tasks such as super-resolution, synthesis and denoising for imaging  \cite{creswell2018generative}, showing better performance comparing to traditional methods when applying abundant training data. Some researchers have also applied this state-of-art method in CT denoising and demonstrated ideal results without complex procedures  \cite{wolterink2017generative,yang2018low}. However, using GAN method for reducing noise level in SPECT images is still under explored \cite{Mok2018initial,Ramon2018initial}.\\
In this paper, we aim to apply and evaluate the use of GAN method in static SPECT image denoising to reduce the injection dose based on 10 simulated patient datasets. \\

\vspace{-15pt}
\section{Method}
\begin{figure}[htbp]
 % Caption and label go in the first argument and the figure contents
 % go in the second argument
\centering
\includegraphics[width=0.6\linewidth]{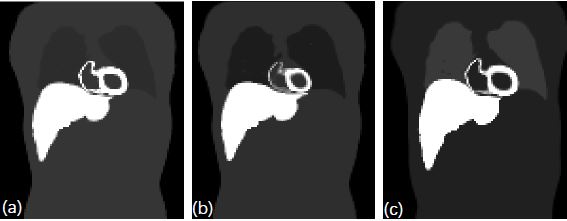}
\vspace{-10pt}
\caption{Three selected XCAT phantoms used in this study.}
\label{fig:Figure 1}
\end{figure}

\textit{Dataset Generation}\vspace{3pt}\\
In order to training and testing proposed network, the 4D Extended Cardiac Torso (XCAT) phantom \cite{segars20104d} was used to simulate 10 male and female patients with different organ sizes and activity uptakes (\figureref{fig:Figure 1}).Nine phantoms were selected for training, while one phantom was chosen for testing. An analytical projector was applied to simulate 120 projections from right anterior oblique to left posterior oblique with two noise levels. The first noise level was based on a standard clinical count rate of 987 MBq injection and 16 min acquisition (low noise) while the other was 1/8 of the previous count rate (high noise). The projections were based on a low energy high resolution collimator, modelling detector-collimator response and attenuation and were then reconstructed by the ordered subset expectation maximization (OS-EM) algorithm with 5 iterations and 6 subsets, using the cine average CT for attenuation correction. The reconstruction matrix size is 128$\times$128$\times$114.
%The routine noise SPECT reconstructed images and corresponding high noisy SPECT reconstructed images were paired from 10 patients’ data, which is shown in (\figureref{fig:Figure 2}). The matrix size of image is 128$\times$128$\times$114. Thus, a total of 114 (number of axial slices obtained from each phantom) $\times$ 10 (number of phantoms) image pairs were generated. \\

\vspace{-02pt}

%\vspace{-20pt}
%\begin{figure}[htbp]
 % Caption and label go in the first argument and the figure contents
 % go in the second argument
%\centering
%\includegraphics[width=0.8\linewidth]{fig2_2.jpg}
%\vspace{-10pt}
%\caption{The diagram of showing 10 patients’ paired SPECT reconstructed images used for GAN.}
%\label{fig:Figure 2}
%\end{figure}
%\vspace{-12pt}

\noindent
\textit{Generative Adversarial Network (GAN)}\vspace{3pt}\\

\vspace{-20pt}
\begin{figure}[htbp]
 % Caption and label go in the first argument and the figure contents
 % go in the second argument
\centering
\includegraphics[width=0.65\linewidth]{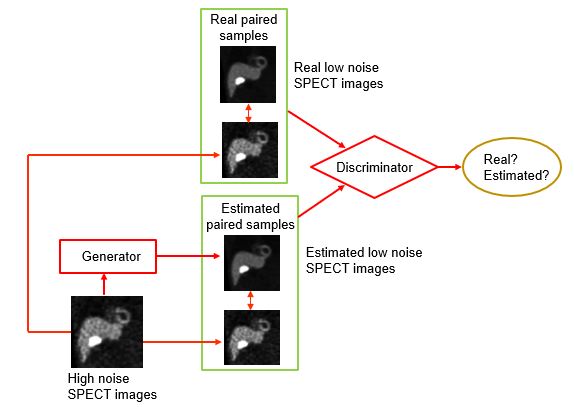}
\vspace{-15pt}
\caption{The conditional GAN used in this work.}
\label{fig:Figure 2}
\end{figure}
\vspace{-2pt}

\begin{figure}[htbp]
 % Caption and label go in the first argument and the figure contents
 % go in the second argument
\centering
\includegraphics[width=0.55\linewidth]{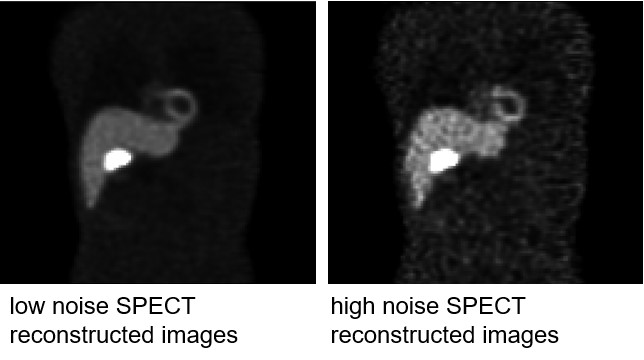}
\vspace{-14pt}
\caption{A sample pair of low noise and high noise SPECT used for training the conditional GAN.}
\label{fig:Figure 3}
\end{figure}

\noindent
Generative Adversarial Network (GAN) is a method of unsupervised learning using two neural networks against each other \cite{goodfellow2014generative}. 
 It consists of a generative network (generator) and a discriminant network (discriminator). The generator takes random sampling from latent space as input, and its output imitates the real samples in the training set. The discriminator aims to distinguish the real sample from the output of the generator. The two networks work against each other and constantly adjust their parameters. The final goal is to make the discriminator unable to discriminate the output of the generator from the real images. Conditional GAN is formed when the input of the original GAN is conditioned with additional information \cite{isola2017image} and is used in this study (\figureref{fig:Figure 2}). The high noise SPECT images were input to the generator while the discriminator compares the generated samples with the “real” samples, i.e., the low noise SPECT images. The calculated loss, i.e., the difference between the generated images and the real samples, would be used for tuning the generator and discriminator simultaneously. This conditional GAN was implemented in Torch and ran on a NVIDIA GeForce GTX 1070 GPU. Both generator and discriminator were optimized by using the Adam optimizer with a learning rate of 0.00001 and 800 training epochs. The total training time was ~2.7 hrs. The high noise and low noise SPECT images of nine patients, i.e., a total of 1026 images (9$\times$114 axial slices) respectively, were paired for training (\figureref{fig:Figure 3}) while 1 patient with high noise SPECT images were tested using the trained conditional GAN.\\

\vspace{-15pt}
\vspace{-18pt}

\begin{comment}
\vspace{-10pt}
\begin{figure}[htbp]
\centering
\begin{minipage}[t]{0.48\textwidth}
\centering
\includegraphics[width=0.85\linewidth]{fig3_04_06.jpg}
\vspace{-14pt}
\caption{The loss curve during training proposed model.}
\label{fig:Figure 3}
\end{minipage}
\begin{minipage}[t]{0.48\textwidth}
\centering
\includegraphics[width=0.95\linewidth]{fig4_04_06_revised.jpg}
\vspace{-14pt}
\caption{A sample set of low noise and high noise images used for training the conditional GAN.}
\label{fig:Figure 4}
\end{minipage}
\end{figure}

\vspace{-16pt}
\end{comment}

\noindent
\textit{Data Post-processing and Analysis}\vspace{3pt}\\
The noise level is measured by the normalized standard deviation (NSD) on a 2D uniform region-of-interest (ROI) with 82 pixels on the liver, in order to compare the results for with and without conditional GAN denoising on the tested SPECT reconstructed images. \\

\vspace{-15pt}
\section{Results and Conclusion}
\begin{figure}[htbp]
 % Caption and label go in the first argument and the figure contents
 % go in the second argument
\centering
\includegraphics[width=0.8\linewidth]{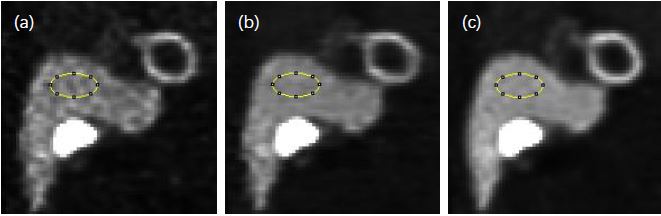}
\vspace{-10pt}
\caption{Sample high noise reconstructed images (a) before using GAN, (b) after using GAN and (c) the original low noise level.}
\label{fig:Figure 4}
\end{figure}

The noise level is substantially reduced in high noise SPECT reconstructed images after using the GAN. The NSD values are 0.1213 and 0.0693 respectively for without and with denoising (\figureref{fig:Figure 4}a and \figureref{fig:Figure 4}b). The NSD value of the low noise SPECT images is 0.0502 (\figureref{fig:Figure 4}c).\\
This proposed method has the potential to decrease the noise level of SPECT images, leading to a reduced injection dose or acquisition time while still maintaining the similar image quality as compared to the original low noise images for clinical diagnosis. Further investigation of this method using the clinical SPECT patients’ datasets are warranted.\\

\vspace{-12pt}
\vspace{-15pt}

% Acknowledgments---Will not appear in anonymized version
\midlacknowledgments{This work was supported by the research grants from the National Natural Science Foundation of China (NSFC), China (81601525), Science and Technology Fund (FDCT) of Macau (114/2016/A3) and University of Macau, Macau (MYRG2016-00091-FST).}

\bibliography{midl-samplebibliography}

\begin{thebibliography}{9}
\providecommand{\natexlab}[1]{#1}
\providecommand{\url}[1]{\texttt{#1}}
\expandafter\ifx\csname urlstyle\endcsname\relax
  \providecommand{\doi}[1]{doi: #1}\else
  \providecommand{\doi}{doi: \begingroup \urlstyle{rm}\Url}\fi

\bibitem[Creswell et~al.(2018)Creswell, White, Dumoulin, Arulkumaran, Sengupta,
  and Bharath]{creswell2018generative}
Antonia Creswell, Tom White, Vincent Dumoulin, Kai Arulkumaran, Biswa Sengupta,
  and Anil~A Bharath.
\newblock Generative adversarial networks: An overview.
\newblock \emph{IEEE Signal Processing Magazine}, 35\penalty0 (1):\penalty0
  53--65, 2018.

\bibitem[Garcia(2012)]{garcia2012physical}
Ernest~V Garcia.
\newblock Physical attributes, limitations, and future potential for pet and
  spect.
\newblock \emph{Journal of Nuclear Cardiology}, 19\penalty0 (1):\penalty0
  19--29, 2012.

\bibitem[Goodfellow et~al.(2014)Goodfellow, Pouget-Abadie, Mirza, Xu,
  Warde-Farley, Ozair, Courville, and Bengio]{goodfellow2014generative}
Ian Goodfellow, Jean Pouget-Abadie, Mehdi Mirza, Bing Xu, David Warde-Farley,
  Sherjil Ozair, Aaron Courville, and Yoshua Bengio.
\newblock Generative adversarial nets.
\newblock In \emph{Advances in neural information processing systems}, pages
  2672--2680, 2014.

\bibitem[Isola et~al.(2017)Isola, Zhu, Zhou, and Efros]{isola2017image}
Phillip Isola, Jun-Yan Zhu, Tinghui Zhou, and Alexei~A Efros.
\newblock Image-to-image translation with conditional adversarial networks.
\newblock In \emph{Proceedings of the IEEE conference on computer vision and
  pattern recognition}, pages 1125--1134, 2017.

\bibitem[Mok et~al.(2018)Mok, Zhang, Cun, Zhang, Pretorius, and
  King]{Mok2018initial}
Greta~SP Mok, Qi~Zhang, Xiaodong Cun, Duo Zhang, P.~Hendrik Pretorius, and
  Michael~A. King.
\newblock Initial investigation of using a generative adversarial network for
  denoising in dual gating myocardial perfusion spect.
\newblock In \emph{2018 IEEE Nuclear Science Symposium and Medical Imaging
  Conference (NSS/MIC)}, pages 1--3. IEEE, 2018.

\bibitem[Ramon et~al.(2018)Ramon, Yang, Pretorius, Johnson, King, and
  Wernick]{Ramon2018initial}
A.~Juan Ramon, Yongyi Yang, P.~Hendrik Pretorius, Karen~L. Johnson, Michael~A.
  King, and Miles~N. Wernick.
\newblock Initial investigation of low-dose spect-mpi via deep learning.
\newblock In \emph{2018 IEEE Nuclear Science Symposium and Medical Imaging
  Conference (NSS/MIC)}, pages 1--3. IEEE, 2018.

\bibitem[Segars et~al.(2010)Segars, Sturgeon, Mendonca, Grimes, and
  Tsui]{segars20104d}
WP~Segars, G~Sturgeon, S~Mendonca, Jason Grimes, and Benjamin~MW Tsui.
\newblock 4d xcat phantom for multimodality imaging research.
\newblock \emph{Medical physics}, 37\penalty0 (9):\penalty0 4902--4915, 2010.

\bibitem[Wolterink et~al.(2017)Wolterink, Leiner, Viergever, and
  I{\v{s}}gum]{wolterink2017generative}
Jelmer~M Wolterink, Tim Leiner, Max~A Viergever, and Ivana I{\v{s}}gum.
\newblock Generative adversarial networks for noise reduction in low-dose ct.
\newblock \emph{IEEE transactions on medical imaging}, 36\penalty0
  (12):\penalty0 2536--2545, 2017.

\bibitem[Yang et~al.(2018)Yang, Yan, Zhang, Yu, Shi, Mou, Kalra, Zhang, Sun,
  and Wang]{yang2018low}
Qingsong Yang, Pingkun Yan, Yanbo Zhang, Hengyong Yu, Yongyi Shi, Xuanqin Mou,
  Mannudeep~K Kalra, Yi~Zhang, Ling Sun, and Ge~Wang.
\newblock Low-dose ct image denoising using a generative adversarial network
  with wasserstein distance and perceptual loss.
\newblock \emph{IEEE transactions on medical imaging}, 37\penalty0
  (6):\penalty0 1348--1357, 2018.

\end{thebibliography}

\end{document}